\documentclass[prl,preprint,showpacs]{revtex4}
\usepackage{graphicx}
\usepackage{amssymb}

\begin{document}

\title{Interplay between shear loading and structural aging in a
physical gel}

\author{O. Ronsin, C. Caroli, T. Baumberger}
\affiliation{INSP, UPMC Univ Paris 06, CNRS UMR 7588
140 rue de Lourmel, 75015 Paris France}

\date{\today}

\begin{abstract}
We show that the aging of the mechanical relaxation of a gelatin gel
exhibits the same scaling phenomenology  as polymer and colloidal
glasses. Besides, gelatin is known to exhibit logarithmic structural
aging (stiffening). We find that stress accelerates this
process. However, this effect is definitely irreducible to a mere
age shift with respect to natural aging. We suggest that it is
interpretable in terms of elastically-aided elementary
(coil$\to$helix) local events whose dynamics gradually slows down as
aging increases geometric frustration.

\end{abstract}


\maketitle

Since glassy materials are out-of-equilibrium metastable systems,
their physical properties slowly evolve with time, a process known as
structural recovery, which gives rise to gradual aging of
thermodynamic quantities. For instance, the specific volume of glassy
polystyrene decreases logarithmically with age, i.e. waiting time
$t_{w}$ after quenching \cite{McKenna}. Besides, the rheological 
response to shear
loading at age $t_{w}$ depends on both the measurement time $t_{w}+t$
and $t_{w}$ itself \cite{Struik}. Creep compliances obey a self-similar scaling
$J(t_{w},t) = \mathcal J(t/t_{w}^\mu )$. In polymer glasses, at low
stress levels, the aging exponent $\mu\lesssim 1$. It decreases at
large stresses approaching yield level. Whether such ``rejuvenation"
is truly equivalent to a shift of the age $t_{w}$ still remains a
matter of debate \cite{McKenna,Rottler, RottlerSGR}. If such is not the case,
as hinted by works of McKenna on polymer glasses \cite{McKenna} and
Viasnoff et al. \cite{ViasnoffPRL, ViasnoffFaraday}
on a colloidal glass, a double question remains: (i) how can this
process be understood in terms of exploration of configurational
space? (ii) to which extent is the answer generic or dependent on the
class of materials?

\begin{figure}[h]
       \centering
       \includegraphics{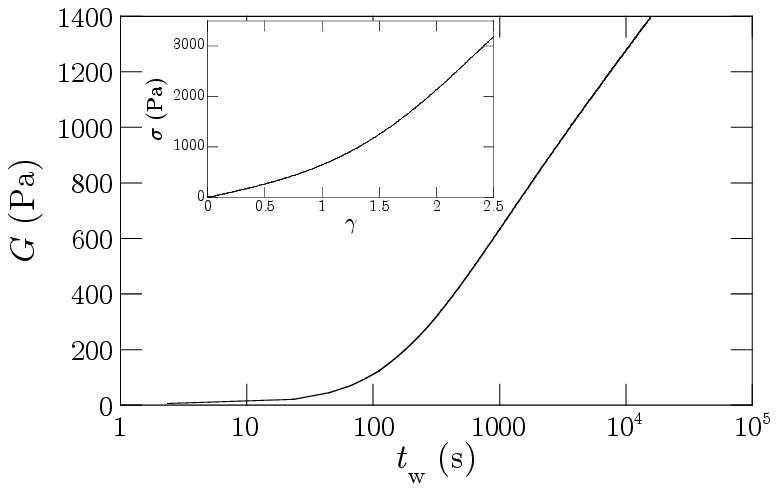}
       \caption{Aging of the small strain shear modulus $G$ measured at $10$ Hz and strain amplitude $10^{-3}$ at temperature $T = 20^\circ$ C.
       Insert : loading curve at age $t_\mathrm{w} = 1000$ s.}
       \label{Fig:Modulus}
\end{figure}

In order to shed further light on these questions, we report here the
results of a study of aging in a gelatin gel.
Gelatin is a physical gel, namely its gelation is thermoreversible
\cite{Nijenhuis}.
The sol state ($T>T_{\rm gel}$) is a solution of single chains of
denaturated collagen in water. Below $T_{\rm gel}$ renaturation of  the native
triple helix structure, stabilized by H-bonds,  becomes
thermodynamically favorable, and chains form a percolating network of
helical segments --- the cross-links (CL) --- connected by single
strand coils. Renaturation is frustrated by strong topological
constraints~ : indeed, since chain length is very long ($\sim
\mu$m), each of them is involved in many CL, hence a large interchain
connectivity. As time after quench increases, after a rapid initial
rise, the gel stiffness  reaches a slow, logarithmic growth regime
(Fig.\ref{Fig:Modulus}) the termination of which has never been
observed \cite{NormandAging}. Several studies
\cite{TerentjevKinetic,Colby} converge towards a common picture~: while,
at early times, structural aging results mainly  from the increase of
the number of CL, in the log regime it is essentially controlled by
CL growth and internal rearrangements. Due to the large
interconnectivity CL growth induces growing internal tensions and
torques on the network-forming coil strands. This is what we call
``increasing geometric frustration". The mechanical relaxation
spectrum consists of two well separated parts \cite{Nijenhuis}~: (i) 
a high-frequency
band, (typically $\omega > 10^{5}$ rad.s$^{-1}$) due to the
viscoelasticity of coil
segments (of length the mesh size $\xi$ typically  $\sim 10$ nm) (ii) 
an ultralow
frequency one, which gives rise to slow creep \cite{RossMurphy} and stress
relaxation \cite{Ferry}.
At intermediate frequencies, the gel is purely elastic and
characterized by its small strain shear modulus $G$.

In summary,  as already shown by
Normand \cite{NormandGlass}, gelatin exhibits a behavior akin to 
glassy dynamics.
Here we investigate in detail the age dependence of mechanical
relaxations and the interplay between structural aging, as measured via
the evolution of $G$, and external loading.

\paragraph{Experimental ---} Samples are prepared by dissolving $5$
wt\% gelatin (300 Bloom, from porcine skin, Sigma) in deionized water at
$80^\circ$C. The gelation temperature is $T_{\rm gel}\simeq
29^\circ$C. The pregel solution is poured into the sand-blasted
cone-plate cell of
a stress-controlled rheometer (Anton Paar, MCR 501), protected
against solvent evaporation by a dodecane rim. Mechanical stability
of the whole rheometer has been improved by enclosing it into a box,
thermalized at $20 \pm 0.1^\circ$C. Sample temperature is first set
at $T_{0} = 50^\circ$C, then ramped at $7.5^\circ$C/min down to the
working temperature $T$. Unless otherwise specified, $T = 20 \pm 0.1^\circ$ C.
We define the onset of gelation, taken to be the
origin of waiting times $t_{w}$, as the time where the loss tangent
$\tan \delta =1$. Reinitialization of the gel history is performed by
reheating up to $T_{0}$, shearing at $\dot\gamma = 1$ s$^{-1}$ for
$200$ s, then repeating the quench.  We have
checked that this protocol ensures that $G$, as well as relaxation
curves, are reproducible to within $1\% $, over 20 cycles at least.
Thanks to this, we are able to probe the evolution of
$G$ along the course of a relaxation in a fully non-perturbative way.
For example, when probing stress relaxation of a gel of age $t_{w}$,
we let it relax for a time $t$, then measure $G(t, t_{w})$ immediately
after fast  unloading. The sample is then reinitialized, and the process is
repeated with a different $t$ value.

The stiffness of the gel is controlled, over a wide strain range,
by the entropic elasticity of single strand coils. A
typical loading curve $\sigma(\gamma)$ is shown on
Fig.\ref{Fig:Modulus} (inset). The
linear range extends up to $\gamma\approx 40\% $, beyond which the
gel strain-hardens. At $\gamma \approx 250\% $, apparent failure is
observed, resulting from wall slip. So, we cannot reach shear melting
(known from fracture studies \cite{FractureNature} to occur for 
stresses $\sim 10^{2}
G$), nor failure of the material itself, and our experiments pertain
to the strongly sub-yield regime.

\begin{figure}[h]
       \centering
       \includegraphics[width=8.6cm]{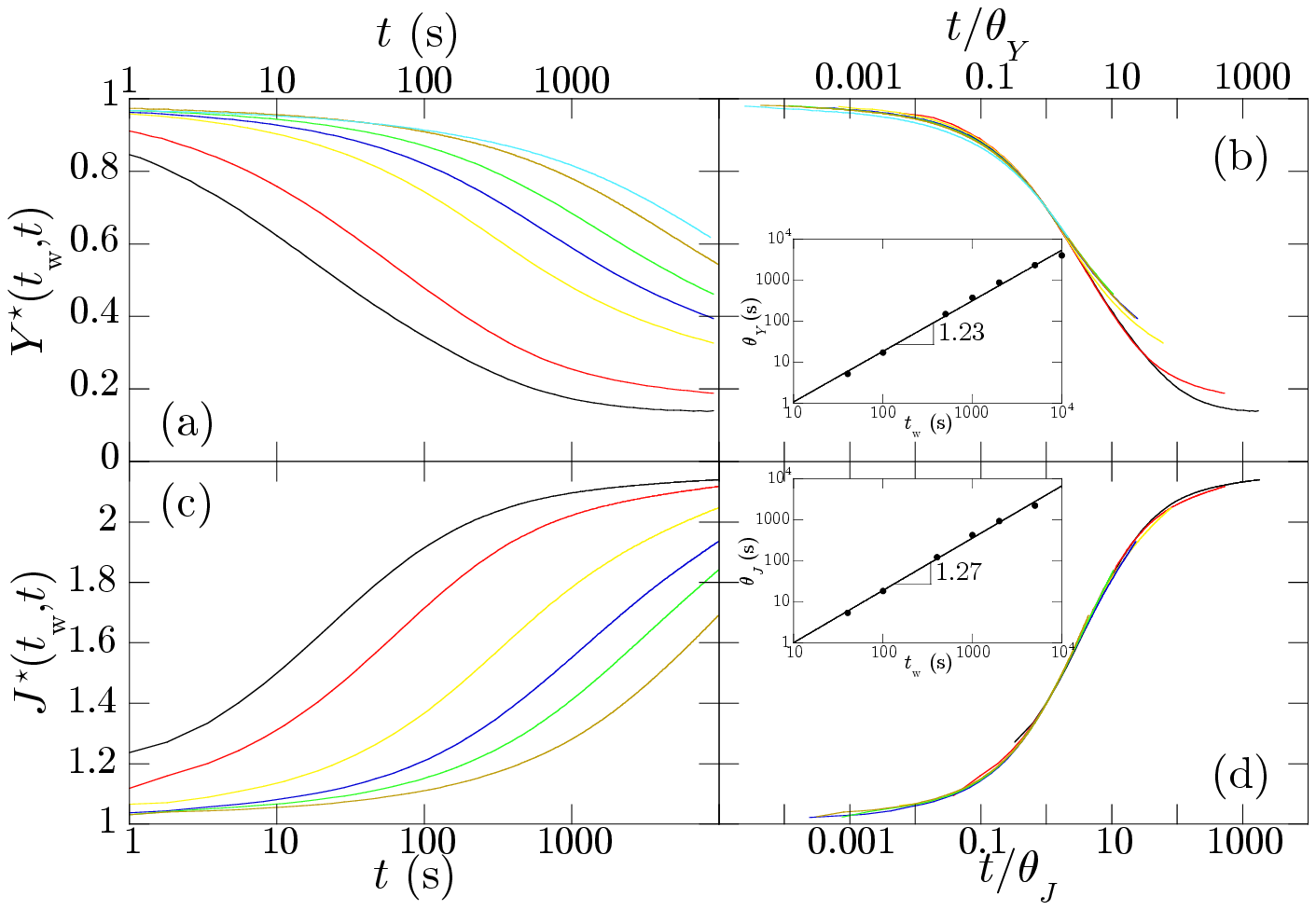}
       \caption{(a) Normalized stress relaxation modulus $Y^\star(t_{w},t) = Y(t_{w},t)/Y(t_{w},0)$ for waiting times, from bottom to top, $t_\mathrm{w} = 40$, $100$, $500$, $10^3$, $2\;10^3$, $5\;10^3$ and $10^4$ s. (b) Same data as (a) plotted \emph{vs.} rescaled time $t/\theta_{Y}(t_\mathrm{w})$. $\theta_{Y}(t_\mathrm{w})$ (see insert) is chosen so that $Y^\star(t_{w},\theta_{Y}) = 0.7$.
       (c) Normalized creep compliance $J^\star(t_{w},t) = J(t_{w},t)/J(t_{w},0)$ for waiting times, from top to bottom, $t_\mathrm{w} = 40$, $100$, $400$, $10^3$, $2\;10^3$ and $5\;10^3$ s. (d) Same data as (c) plotted \emph{vs.} rescaled time $t/\theta_{J}(t_\mathrm{w})$. $\theta_{J}(t_\mathrm{w})$ (see insert) is chosen so that $J^\star(t_{w},\theta_{J}) = 1.4$.}
       \label{Fig:Scalings}
\end{figure}

The evolution with waiting time of
the stress relaxation (SR) and creep (Cr) responses is shown on
Figs.\ref{Fig:Scalings}.a,c which correspond to loadings in the
linear elastic range. In this regime, we find that the responses are themselves
linear, characterized by the creep compliance $J(t_{w},t)$ and the
stress relaxation modulus $Y(t_{w},t)$.

Figs.\ref{Fig:Scalings}.b,d show that both quantities can be quite
satisfactorily collapsed by rescaling time by $t_{w}$-dependent
factors $\theta_{Y}$ and $\theta_{J}$ which we find (see insets) to
obey, over more than 2 decades, power law scalings with the same
exponent~:
\begin{equation}
       \theta_{Y,J}\sim t_{w}^\mu\,\,\,\, {\rm with}\,\, \mu = 1.25\pm0.02
       \label{}
\end{equation}

So, in the strongly sub-yield regime, the gel system exhibits the
same rheological aging phenomenology as glasses. Yet, while most
glassy materials are of the sub-aging type ($\mu <1$) \cite{McKenna,
RottlerSGR}, gelatin turns
out to be hyper-aging. However, closer inspection of
Fig.\ref{Fig:Scalings}.b reveals a noticeable splay of the scaled
curves beyond $t/\theta_{Y}\approx 1$. Moreover, a trend towards
saturation of the shear stress at a finite level is clearly visible
(see also Fig.\ref{Fig:Scalings}.a). Creep curves
for young gels ($t_{w}\leq 100$ s) exhibit, after a quasi-logarithmic
intermediate regime,  a similar trend towards
strain saturation. How can we understand this previously unreported
behavior? It is natural here to decompose the total strain $\gamma$
as the sum of an elastic component and a plastic one, i.e.~:
\begin{equation}
       \gamma = \frac{\sigma}{G} + \gamma_{\rm pl}
       \label{Eq:Gamma}
\end{equation}
It is then clear that the results of standard relaxation experiments
at constant $\sigma$ or $\gamma$ mix information about the aging of
structure and of flow properties.
Indeed, although its dynamics may be affected by the mechanical
perturbation, structural aging is certainly at work, so that for
large $t\gg t_{w}$, the value of $G$ is certainly not coded simply by
the initial age $t_{w}$ but, rather, by the true one $(t_{w}+t)$.
Hence the limited validity of the above scaling and the need for
characterizing structural aging under mechanical perturbations.

\begin{figure}[h]
       \centering
       \includegraphics{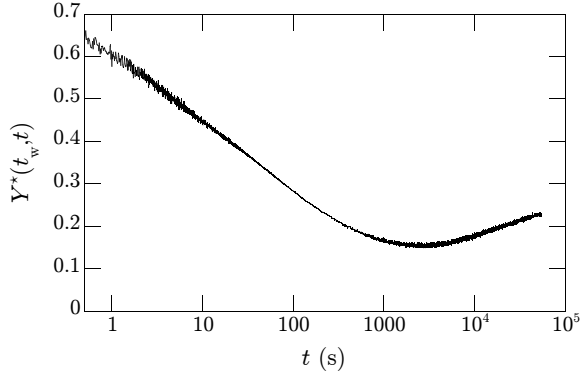}
       \caption{Normalized stress relaxation modulus \emph{vs.} time for a gel of age $t_\mathrm{w} \simeq 10$ s at temperature $T = 10^\circ$ C.}
       \label{Fig:Buildup}
\end{figure}

We systematically measured the elastic modulus $\tilde G(t_{w}, t)$
along the course of SR experiments performed at various $t_{w}$ and
$\gamma$-levels in the linear elastic regime.
We find  that, within experimental accuracy, in all cases $\tilde
G(t_{w},t) = G(t_{w}+t)$, i.e. under such conditions, natural
structural aging
is unperturbed, and is likely to control the long term dynamics. If
so, $\dot\sigma = \dot G\sigma/G -\dot\gamma_{\rm pl}G$ might change
sign. We have indeed evidenced such a late ($t \gtrsim 300\,t_{w}$)
stress build-up regime (see Fig.\ref{Fig:Buildup}) by
taking  advantage of the fact that cooling from
$20$ to $10^\circ$C increases the natural aging log-rate  by a factor
of 2.5 \cite{Nijenhuis}. We suggest that this behavior might be the 
SR-analog of the
reversal between early creep and late strain recovery observed by
Cloitre {\it et al.} (see \cite{Cloitre}, figure 4).

SR is certainly poorly suited to reveal a possible shear-sensitivity
of structural aging since, even for large imposed strains, $\sigma$
only remains noticeable for a limited time. We have therefore
measured $\tilde G(t_{w}, t)$ in the Cr configuration. Again, no
departure from natural aging is measurable, for $t$ values up to
$500\,t_{w}$, up to stress levels $\sigma = G(t_{w})$ corresponding to
{\it initial} strains of $100\% $. However, since the network gradually
stiffens, the dimensionless strength $\sigma/\tilde G$ decreases, and
stress itself is probably not a good control parameter. As an attempt to
circumvent this drawback, we have devised ``assisted creep" (ACr)
experiments in which we apply to the sample an increasing stress
$\sigma = \gamma_{0}G(t_{w}+t)$ with $G$ the value for natural aging.
This protocol is meant to work, as far as possible, at constant
elastic strain $\gamma_{\rm el} = \gamma_{0}$.

\begin{figure}[h]
       \centering
       \includegraphics{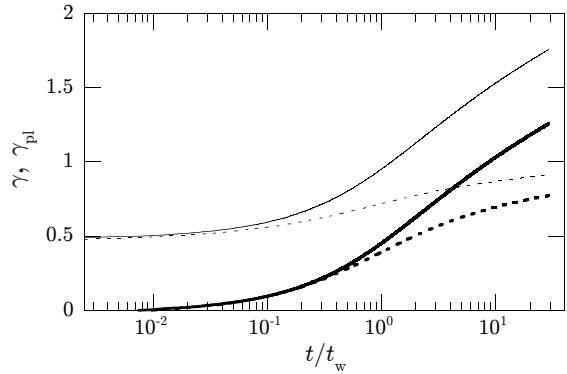}
       \caption{Assisted (full curves) and standard (dashed curves) creep total (thin lines) and plastic (thick lines) strain responses for a gel of age $t_\mathrm{w} = 400$ s. At the end of the ACr run, $\gamma_\mathrm{el} = 0.97\gamma_0$.}
       \label{Fig:ACr}
\end{figure}

\begin{figure}[h]
       \centering
       \includegraphics{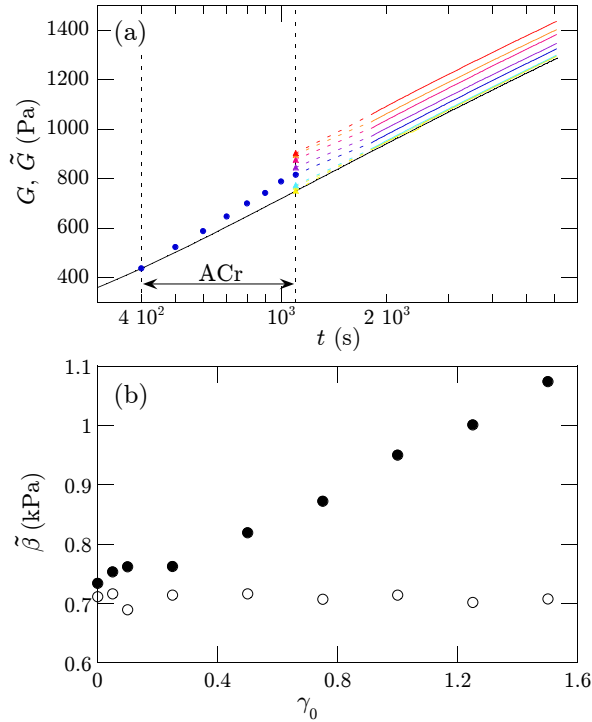}
       \caption{(a) Dots: aging of the shear modulus $\tilde{G}$ (t) during an ACr experiment with $\gamma_0=0.5$. Each datum is obtained by the unloading slope of a run stopped at $t$ (see text). Lines : aging after strain recovery following an ACr under $\gamma_0$ values, from bottom to top : $\gamma_0 = 0.05$, $0.1$, $0.25$, $0.5$, $0.75$, $1.0$, $1.25$ and $1.5$. Triangles show unloading slopes at the end of the ACr phase. Thick line : natural aging $G(t)$. (b) Aging log-slope $\tilde{\beta}$ during (full dots) and after (empty dots) assisted creep plotted \emph{vs.} $\gamma_0$.}
       \label{Fig:OverAging}
\end{figure}

Fig.\ref{Fig:ACr} shows the results of a standard and an assisted
creep experiments performed on equally-aged gels. The ACr enhancement
of the flow dynamics $\gamma_{\rm pl}(t)$ (eq. (\ref{Eq:Gamma})) in
the intermediate, quasi-logarithmic regime is
spectacular. We have performed a set of ACr runs in which various
values  of $\gamma_{0}$ are applied to a gel of age $t_{w}=400$~s for a
duration  $\Delta t = 700$~s after which we unload to zero shear
stress. The small strain modulus, measured during and after the
mechanical perturbation, is shown on Fig.\ref{Fig:OverAging}. The
effect of external loading is now unmistakable: (i) under a finite
$\gamma_{0}$, structural aging is accelerated. The larger
$\gamma_{0}$, the larger the log-slope $\tilde \beta = d\tilde
G/d(\log t)$ (see inset) (ii) after unloading, $\tilde\beta$ recovers
its $\sigma =0$ value ($\beta = 700$ Pa) and, for $t>t_{w}+\Delta t$,
the only memory of the loading episode kept by the system consists in
a rigid shift of $\tilde G(t)$ with respect to its ``natural" value
$G(t)$. So, although loading induces accelerated structural
strengthening, this effect is by no means equivalent to a mere
forward shift of the ``natural age". We thus confirm the conclusion
of McKenna and Viasnoff et al. that mechanical perturbations of
slow glass-like relaxation cannot truly be termed overaging (nor,
alternately, rejuvenation).

\paragraph{Discussion} --- One step further, the above set of
results leads us to propose the following tentative picture for
physical aging in gelatin. As proved by Djabourov et al.
\cite{Djabourov}, $G$ and
the helix fraction obey a one-to-one relation. We focus here on
the logarithmic regime where  nucleation of new CL
is negligible, so that natural aging is ruled by the growth of
preexisting ones \cite{Colby}, at the expense of the connecting coils, the
stiffness of which controls the gel modulus $G$. So, due to
solvent incompressibility, the average mesh size remains
quasi-constant and $G$ grows. The formation of a new unit helix
segment  (h) can be pictured as a H-bonding reaction involving one
monomer from each of the three coils (c) emanating from the CL end.
In order for the reaction (c $\to$ h) to proceed, these three
monomers must
``meet" in the proper positional and orientational
configuration. This topological constraint can be depicted as an
entropic activation barrier separating the c-state from the
energetically favorable h-one ($F_{h}<F_{c}$).

As shown by
Kutter and Terentjev \cite{TerentjevCoilHelix}, as the coil length 
decreases under these
conditions, $F_{c}$ itself decreases. At the same time, coil
shortening induces a thinning of the entropy supply, and thus
an increase of the barrier height $F_{b}$.
In the spirit of the analysis, by Knoll et al. \cite{Knoll} of the
relaxation kinentics of nano-indents in a polymer glass, we make the
schematic assumption that (i) the barrier height $\mathcal E =
F_{b}-F_{c}$ increases linearly with the average helix fraction
$\chi$ (i.e. with CL length)  (ii) $\chi$ evolves with an Arrhenius
dynamics $\dot\chi = \tau^{-1}\exp[-\epsilon\chi/k_{B}T]$ with
$\epsilon = d\mathcal E/d\chi$ the ``sensitivity to frustration".
This highly schematic model predicts that the frustration-induced
slowing down of the c$\to$h reaction results in a logarithmic
dynamics where the slope $d\chi /d\ln t = k_{B}T/\epsilon$ is
controlled by the sensitivity parameter $\epsilon$.

When an elastic strain $\gamma_{\rm el}$ is imposed, the elastic
energy is stored in the compliant coil, leading to an upward
shift $\Delta F_{c}$ and thus to  acceleration of
aging. The barrier free energy is also shifted, though in an
anisotropic fashion: along the stretching (resp. compressive)
principal direction, coil entropic wandering are restricted
(resp. facilitated) and $\Delta F_{b} >0$ (resp. $<0$). So CLs grow
faster along the compressed direction than along the stretched
one. When unloading, the corresponding  relative coil shortening
leads to a remanent, plastic strain. We believe creep to be due
to this texturing, differential rate-of-growth effect, rather than
to CL ``melting" under stress, as initially proposed by Ferry.
Indeed, on the one hand, the probability of the h$\to$c reaction is
negligibly small with respect to that of the c$\to$h one, since
the binding energy $F_{c}-F_{h}$ is on the order of that of 3
H-bonds ($\sim
0.3$ eV $\gg k_{B} T$) \cite{Activecrack}. On the other hand,
melting would be
contradictory with the observed stress-induced acceleration of
stiffening.
Finally, once the ``mechanically-overaged" system is unloaded,
the elastic shift is suppressed, and the CL growth dynamics
returns to the natural one, the memory of the perturbation being
encoded in the height reached by the entropic barrier at the end
of the loading phase. Hence the recovery of the log-slope
$\beta$.

In summary, the experimental results of a coupled study of
mechanical relaxations and of the stress-induced acceleration of
the stiffening dynamics leads us to propose that physical aging
in gelatin gels can be described in terms of {\it local}
irreversible events. Namely, we picture CL growth as the
sequential formation of unit helix segments, via activation over
an entropic barrier whose height increases with the degree of
completion of relaxation towards thermodynamic equilibrium. That
is, this barrier height appears, roughly speaking, as an
``order parameter" into which geometric frustration effects are
lumped. Whether such a simple,  local description  of aging
could make sense in glassy systems remains an open issue.
If so, it would mean that aging would be controlled by the
growth of the instability threshold of the local cluster rearrangements,
hence by the relaxation of the
average free-volume. This is precisely the subject of an
active ongoing debate. We believe that experimental studies such as that
of specific volume relaxation {\it after} an ``implosion"
episode of the type reported by McKenna could bring valuable
insight into this issue.

\begin{acknowledgments}

\end{acknowledgments}


\begin{thebibliography}{100}
\bibitem{McKenna} G. B. McKenna, J. Phys. Condens. Matter {\bf 15},
S737 (2003).

\bibitem{Struik} L. C. E. Struik \emph{Physical Aging in Polymers and 
Other Amorphous Materials} (1978) (Amsterdam: Elsevier).

\bibitem{Rottler} J. R\"ottler and M. Warren, Eur. Phys. J. Special 
Topics {\bf 161}, 55 (2008).

\bibitem{RottlerSGR} M. Warren and J. R\"ottler, Phys. Rev. E {\bf 
78}, 041502 (2008) and references therein.

\bibitem{ViasnoffPRL} V. Viasnoff and F. Lequeux, Phys. Rev. Lett. 
{\bf 89}, 065701 (2002).

\bibitem{ViasnoffFaraday} V. Viasnoff, S. Jurine and F. Lequeux, 
Faraday Discussion {\bf 123}, 253 (2003).

\bibitem{Nijenhuis} K.~te~Nijenhuis, Adv. Polymer Science \textbf{130} (1997).

\bibitem{NormandAging} V. Normand, S. Muller, J.-C. Ravey and A. 
Parker, Macromolecules {\bf 33}, 1063 (2000).

\bibitem{TerentjevKinetic} J. L. Gornall and E. M. Terentjev, Phys. 
Rev. E {\bf 77}, 031908 (2008).

\bibitem{Colby} L. Guo, R. H. Colby, C. P. Lusigan and A. M. Howe, 
Macromolecules {\bf 36}, 10009 (2003).

\bibitem{RossMurphy} P. M. Gilsenan and S. B. Ross-Murphy, Int. J. 
Biol. Macromol. {\bf 29}, 53 (2001).

\bibitem{Ferry} M. Miller, J. D. Ferry, F. W. Schremp and J. E. 
Eldridge, J. Phys. Chem. {\bf 55}, 1387 (1951).

\bibitem{NormandGlass} V. Normand and A. Parker, Proceedings of the 
3rd International Symposium on Food Rheology and Structure, ETH 
Z\"urich, 185 (2003).

\bibitem{FractureNature} T. Baumberger, C. Caroli, D. Martina, Nature 
Materials, {\bf 5}, 552 (2006).

\bibitem{Cloitre} M. Cloitre, R. Borrega and L. Leibler, Phys. Rev. 
Lett. {\bf 85}, 4819 (2000).

\bibitem{Djabourov} C. Joly-Duhamel, D. Hellio, A. Ajdari and M. 
Djabourov, Langmuir {\bf 18}, 7158 (2002).

\bibitem{TerentjevCoilHelix} S. Kutter and E. M. Terentjev, Eur. 
Phys. J. E {\bf 8}, 539 (2002).

\bibitem{Knoll} A. Knoll, D. Wiesmann, B. Gotsmann, and U. Duerig, 
Phys. Rev. Lett {\bf 102}, 117801 (2009).

\bibitem{Activecrack} T. Baumberger and O. Ronsin, J. Chem. Phys. 
{\bf 130} 061102 (2009).

\end{thebibliography}
\end{document}